# Requirements Engineering of a Web Portal using Organizational Semiotics Artifacts and Participatory Practices

Flávia Linhalis Arantes[1]

[1]Nucleus of Informatics Applied to Education (NIED)
State University of Campinas (UNICAMP), Campinas, Brazil
`farantes@unicamp.br`

## ABSTRACT

*The requirements of software are key elements that contribute to the quality and users satisfaction of the final system. In this work, Requirements Engineering (RE) of web sites is presented using an organizational semiotics perspective. They are shown as being part of an organization, with particular practices, rules and views considering stakeholders several differences and opinions. The main contribution of this paper is to relate an experience, from elicitation to validation, showing how organizational semiotics artifacts were exploited in a collaborative and participatory way to RE of a web portal. A case study is described in order to demonstrate the feasibility of using such artifacts to RE when we think about the system as being part of a social organization.*

## 1. INTRODUCTION

One of the measures of success of a software system is concerned about its adequacy to the purpose for which it was intended. According to Baranauskas and colleagues [1], "*Requirements Engineering (RE) is the process of discovering the purpose for which a software system is intended by identifying stakeholders and their needs, and documenting these in a form that is suitable to analysis, communication, and implementation*". It is usual to have conflicts and different views of the tasks by the stakeholders, because it depends on the perspectives of their tasks in the system and work environment. A semiotic-based method for RE was proposed by Baranauskas and colleagues [2, 3]. The method is based on Organizational Semiotics (OS) [4, 5], which studies organizations considering social, political, cultural and ethical issues involved in understanding the design problem.

The Participatory Design (PD) proposes stakeholders participation during the life cycle of a technological product [6]. The users can actively contribute to reflect their perspectives, needs and expectations during the design and development of the system [7].

The methodological conception of the case study presented in this paper considers the technology in a social context, with methods and techniques from OS and PD, which was coined "semio-participatory" methodology by Baranauskas [8]. One of the objectives of the methodology is to promote the active participation of the stakeholders in the process of solution design, implementation and evaluation.

In this work, the semio-participatory methodology [8] is applied to the RE of a web portal. The case study presented in this paper allows us to observe the contribution of OS when we consider the software system as being part of a social organization. The main contribution of this paper is to relate an experience, showing a web portal case study that joins participatory practices and OS artifacts applied to RE – i.e., elicitation, analysis/negotiation, specification, and validation.

This paper is organized as follows: section 2 presents some related work. Section 3 presents OS artifacts and how they that can be applied to RE. Section 4 describes web information systems from OS perspective, giving background to the fact that web information systems design should be contextualized in a social perspective. Section 5 presents a case study, showing how the semio-participatory methodology were applied to the RE phases of a web site. Sections 6 and 7, respectively, present discussion and conclusion about the case study, highlighting the semio-participatory methodology contributions.

## 2. RELATED WORK

The methodology adopted in this research situates the technology in a social context, with methods and techniques from OS and PD – that is called semio-participatory methodology [8].
The use of OS artifacts in conjunction with PD techniques has been explored in previous research work in Science Computing. Baranauskas and colleagues [2] used a semiotic approach to guide the requirements elicitation process. Bonacin and colleagues [9] used OS as a guide to develop a service-based architecture for e-Government. Simoni [10] and colleagues adopted a social process to information systems development, showing a semio-participatory case study. Melo [11] used semio-participatory techniques to the design of inclusive information systems. These works, among others, have contributed to the definition of the semio-participatory methodology approach, adopted in the case study presented in this paper.

Currently, the semio-participatory methodology is being explored to guide the integration of educational laptops in a Brazilian school. The focus is the involvement of the school community to build a participatory model for the educational laptop insertion in the school practices, and to analyze its impacts outside the school. The semio-participatory methodology is helping to understand how the technology is situated in the school social context. Results can be found at [17, 18].

## 3. OS ARTIFACTS APPLIED TO RE

Organizational Semiotics studies organizations based on semiotics concepts and methods. This study understands that any organized behavior is affected through and governed by the communication and interpretation of signs by people [4, 5]. The aims of OS study are to find new and insightful ways of analysing, describing and explaining organizations.

The OS methods and artifacts have been reviewed and used in previous work to perform several tasks related to RE [2, 9, 10]. In the following subsections, some OS artifacts that can be used to support RE phases are described.

### 3.1. OS Onion and Stakeholders Analysis

The semiotic onion (or OS onion) is a conceptual artifact that represents the computational system using different layers of meaning. Each layer has to be considered in the system analysis and design. The relationship between the semiotic onion conceptual layers of an information system is illustrated in Figure 1(a) [4].
As illustrated in Figure 1(a), at the informal system layer there is a sub-culture in which meanings are established, intentions are understood, beliefs are formed and commitments with responsibilities are made. In the formal system layer, form and rule replace meaning and intention. Finally, in the technical layer, part of the formal system is automated by a computer system. The informal level contains the formal one that, by its turn, contains the technical level. Changes in some level have impact in others.

The OS onion allows investigating and identifying the stakeholders that have influences or interests in the information system development. Firstly, the design problem has to be clarified and the stakeholders should be identified and organized in the OS onion layers [2, 9]. Figure 1(b) illustrates the semiotic onion adapted to guide the design problem clarification, stakeholders identification and analysis.

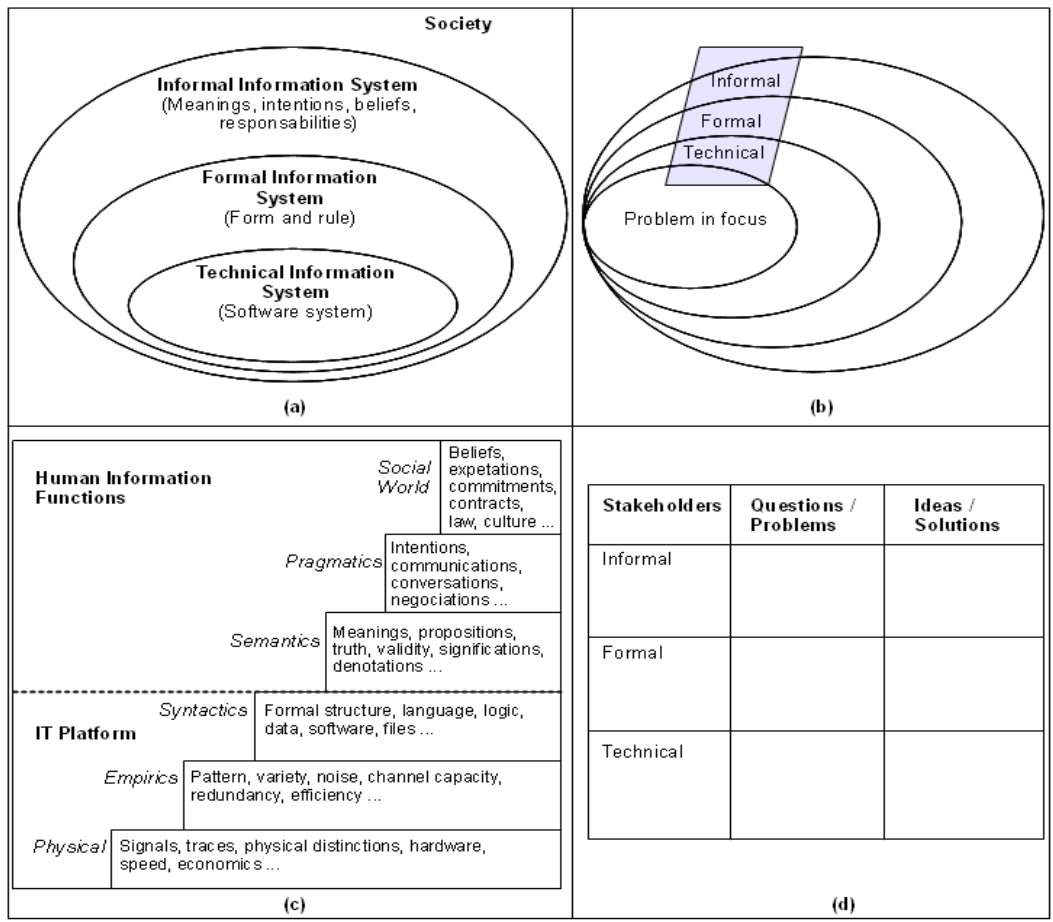

Figure 1. (a) Organizational Semiotics Onion [4]. (b) Organizational Semiotics onion adapted to problem clarification and stakeholders analysis. (c) The Semiotic Ladder [4]. (d) Evaluation frame structure.

## 3.2. Semiotic Ladder

The semiotic ladder (Figure 1(c)) is an artifact that supports a semiotic view of information systems requirements considering the different aspects of information that are organized in different levels of information. This artifact expresses the idea of constructing the system in each level from the ones identified in lower levels [12].

The semiotic ladder has been used to support information system analysis in several phases of software life cycle [2, 13]. The artifact offers the opportunity of analysing semantic, pragmatic and social levels of relationship, while traditional methodologies emphasize mostly technical issues.

## 3.3. Evaluation Frame

Evaluation frames are used to analyze the current system and understand what is expected from the new one. This artifact allows to identify and organize interests, ideas, questions and problems identified by each stakeholder category [2, 9, 10]. Figure 1(d) illustrates an example of the evaluation frame structure.

## 3.4. Ontology Model

The ontology model shapes a context that involves concepts and words used in the domain of a specific problem. It is a semantic model of the problem domain.

An ontology model can be used to represent user's requirements in a formal and precise model [4]. The required system functions are specified in the model, which describes a view of responsible agents in the focal business domain and their actions or patterns of behavior called "affordances" [14]. It is a process of conceptualization of a business organization, in which the organizational behavior and the language used to express the problem are analyzed and captured in the ontology model. The primary focus of the formalization is on the agents in action. The agents and their patterns of behavior (affordances) have a graphical representation, which includes [4]:

- Agent (graphically represented as an ellipse): Actors who build and interact with the reality.
- Affordance (rectangle): Semantic primitive representing possible patterns of agent actions or behaviors.
- Ontology Relation (line): Define the limit or period of existence of an affordance related to the agent that holds it. The antecedent in the relation is represented on the left and the dependent on the right.
- Determiner (preceded by #): Invariant property that distinguish one instance from others.
- Role (half circle): An agent can have a particular role when he or she is involved in relations and actions.
- Whole-part Relationship (line with a dot): Define a possible subdivision of an agent, represented from the left (whole) to the right (part), according to the ontological dependence.
- Generic-specific Relationship (box): specifies whether agents of affordances possess shared properties.

The design problem description and the requirements elicitation performed by users can collaborate to define semantic units that may indicate agents, affordances and their relationships to formalize and model the problem in focus. In section 5.4 an ontology model example is described.

## 4. WEB INFORMATION SYSTEMS WITH AN OS PERSPECTIVE

According to Melo [11], the semiotic onion and the semiotic ladder can support information systems analysis to web information system in general. If we consider the web as an example of technical information system, then the global society could be seen as the informal information system. The norms to rule their actions (development, maintenance, for example) would be part of the formal information system.

If we take an institutional portal as example, then the informal layer of the onion would be represented by the institution that owns the portal, and by other groups interested on its activities. In the formal level, the focus would be in the norms, for example, work procedures, privacy policies, legislation, internal regulation, etc.

The following items describe how the semiotic ladder can be used to support web information systems analysis [11]:

- Physics: hardware infra-structure particularly related to the problem in focus. For example: servers, transmission channels, client computers, etc.
- Empirics: bandwidth, network protocols, data transmission, access efficiency, throughput, etc.
- Syntactics: relates to development frameworks, programming and markup languages used in the web (PHP, ASP, HTML, CSS, etc), W3C recommendations, file systems, documentation, browsers, and so on.
- Semantics: this layer can support different point of views. For example, (1) who are the users, their needs and interests; (2) the meaning of interface elements; etc.
- Pragmatics: established communication, adopted procedures, users intentions, etc.
- Social World: values, expectations, cultural influences, beliefs, compromises, legislation, and so on.

## 5. A WEB PORTAL CASE STUDY

The Nucleus of Informatics Applied to Education (NIED) web portal is the object of this case study. NIED is a research nucleus, which aggregates researchers and projects from several areas, with focus on informatics in education. Its web portal is expected to highlight the nucleus team, projects and advances.

NIED's current web portal was contracted as a third party service. There was no stakeholders participation during the design process. This problem leaded to dissatisfaction and frustration when the portal was delivered.

The portal is now under a (re)design process. Participatory practices [7] are being considered and supported by OS artifacts. Figure 2 illustrates the semio-participatory methodology [8] adopted for the portal RE process – from stakeholders identification to requirements validation. In the following subsections, the Requirements Engineering phases applied to NIED's portal design are described, together with the OS artifacts adopted in four participatory workshops.

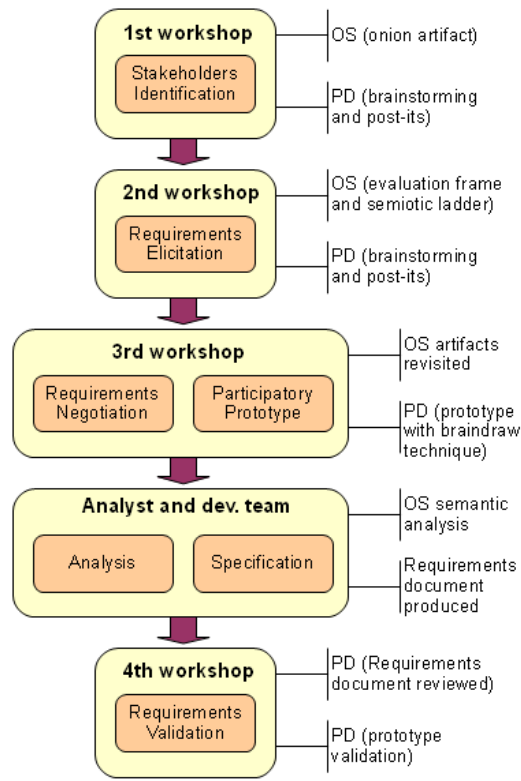

Figure 2. Methodology outlined for the portal Requirement Engineering process – from stakeholders identification to requirements validation.

## 5.1. Stakeholders Identification

The clarification of the design problem and the identification of the stakeholders were the first steps to requirements elicitation. The identified stakeholders were about to take part in the participatory practices. A brief participatory workshop was conducted with NIED's board members in order to perform these two activities.

The OS onion artifact, presented in Figure 1(b), was used to support this activity. The first activity of the workshop was to expose the design problem as being "*the development of a web portal to meet NIED's interdisciplinary nature and to highlight the nucleus projects and advances*". The workshop mediator explained the OS onion artifact to the participants. During a brainstorming, the stakeholders were identified and classified in the semiotic onion layers. Figure 3 summarizes the results.

According to the Figure, in the inner layer, the problem of design (NIED's web portal) is represented. In the technical layer, part of the formal system is automated – so technical stakeholders were identified. In the formal layer, norms, work procedures and internal rules are established, so people who rule work procedures at NIED were identified. In the informal level, social groups interested on NIED's activities were identified.

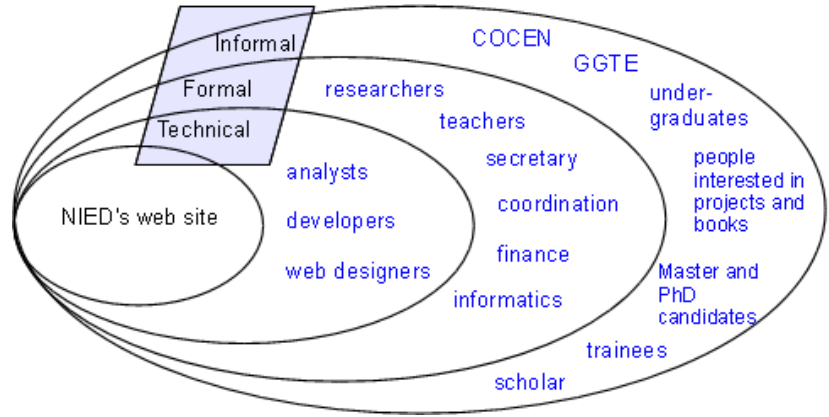

Figure 3. OS onion applied to NIED's web portal stakeholders identification.

## 5.2. Requirements Elicitation

Once the stakeholders were identified, another semio-participatory workshop were conducted with focus on requirements elicitation. At least one person of the following stakeholders groups took part in the workshop: developers, analysts, web designers, researchers, secretary, informatics, undergraduates, PhD candidates, scholars, trainees, COCEN and GGTE.

In the second workshop, the mediator firstly revisited the OS onion artifact, filled in the first workshop. The design problem was also reviewed, as some participants were not present in the first workshop because they were identified as stakeholders in that occasion.

The workshop mediator presented two artifacts that would be used to requirements elicitation: the evaluation frame (Figure 1(d)) and the semiotic ladder (Figure 1(c)).

With support of the evaluation frame artifact, the mediator asked the participants to identify questions, problems, ideas and solutions for the portal and write them in post-its. The stakeholders of each layer (informal, formal and technical) were all together in this activity. After writing in the post-its, each one exposed to the group his/her questions and ideas to NIED's portal, according to the vision they had of the system. In order to organize and register the users contributions, an evaluation frame poster was filled with the post-its by the mediator.

User participation in different semiotic levels led to heterogeneous ideas of the system. When these ideas were exposed to the group, they were enriched in a participatory manner. The items below summarize the evaluation frame main results, showing contributions to the informal, formal and technical layers, respectively. The complete evaluation frame results are shown in Tables 1 to 3.

- **Informal**: The main contributions to the informal level were related to keep the portal up to date, to show NIED's history (including projects and people), to integrate NIED's community, to make internationalization, to make the portal more interesting and interactive (using blogs and community contributions).
- **Formal**: In the formal level, the requirements were related to buy books (electronic commerce), to create an Intranet to administrative subjects, to customize some common areas in the portal, to design simple administration interfaces, to integrate the portal with NIED's research projects, to make a digital library, and to allow multiple users to edit events and news.

- **Technical**: The technical level contributions were mainly related to make the portal accessible (at least to visual impaired users), to adopt RSS and/or social networks, to make the site map, and to access the portal using mobile devices (such as PDAs).

Table 1. Evaluation frame results for informal layer.

| Questions and Problems | Ideas and Solutions |
|---|---|
| 1) The projects in the portal are usually outdated. | 1) Someone have to cover the project's upgrade. |
| 2) Tell the story of NIED, including projects and people. | 2) Make a time line, indicating projects, participants, and important events. |
| 3) More integration of NIED's community (students, researchers, teachers, ...). | 3) Create NIED's community as a social network and access it from the portal. |
| 4) NIED's web portal internationalization | 4) Use automatic translation (like google translate) |
| 5) In the current portal, it is difficult to find NIED's address and location. | 5) Put these information in the main page. |
| 6) Make the portal more interesting. | 6) Increase the visibility of projects with multimedia. |
| 7) Make the portal more interactive | 7) Use blogs |

Table 2. Evaluation frame results for formal layer.

| Questions and Problems | Ideas and Solutions |
|---|---|
| 1) Enable users to buy books. | 1) Implement electronic commerce. |
| 2) Lack of administrative resources in the current portal. | 2) Create an Intranet to administrative subjects. |
| 3) Problems with meeting schedule in NIED's physical space. | 3) Implement scheduling to NIED's meeting room. |
| 4) Customize some common areas in the portal. | 4) Make a minimum template to projects and personal home pages. |
| 5) Problems to keep private areas up to date. | 5) Design simpler administration interfaces to encourage users to update their areas. |
| 6) More integration with NIED's research projects. | 6) Allow access to projects' web pages according to user permissions. |
| 7) More integration with NIED's research area. | 7) Allow community contributions related to projects research focus. |
| 8) It is difficult to keep the portal up to date with events and news. | 8) Decentralize – allow multiple users to edit this area. |
| 9) It is difficult to keep productions (books, memos, papers) up to date. | 9) Decentralize – each researcher responsible for his/her projects. |
| 10) Ease access to NIED's books. | 10) Make a digital library. |

Table 3. Evaluation frame results for technical layer.

| Questions and Problems | Ideas and Solutions |
|---|---|
| 1) Make the portal accessible (at least to visual impaired users). | 1) Adopt W3C WAI (Web Accessibility Initiative). |
| 2) See NIED's news without entering the portal. | 2) Use RSS and/or social networks. |
| 3) It is difficult to find information in the current portal. | 3) Make the site map. |
| 4) Access the portal using other platforms. | 4) Consider mobile devices (such as PDAs). |

The semiotic ladder (Figure 1(c)) was also adopted to support requirements elicitation (mainly non-functional requirements). With this artifact some human aspects of information related to social world (compromises), pragmatics (intentions) and semantics (meanings) were identified. Some compromises registered in the semiotic ladder were related to information security and availability, compliance with legislation (norms, W3C recommendations, accessibility laws), different browsers compatibility, and keeping the portal up to date.

The intentions observed about the stakeholders needs and interests were related to ease the content management, to improve the portal visibility, to improve NIED's projects visibility, and to make the portal more intuitive and dynamic.

The observations about meaning focused on usability, accessibility, internationalization, understandable interfaces, conceptualization of the interface by the stakeholders, and the use of timeline metaphor.

After the workshop, the semiotic ladder were reviewed and updated mainly with infrastructure aspects. The complete semiotic ladder results are shown in Figure 4.

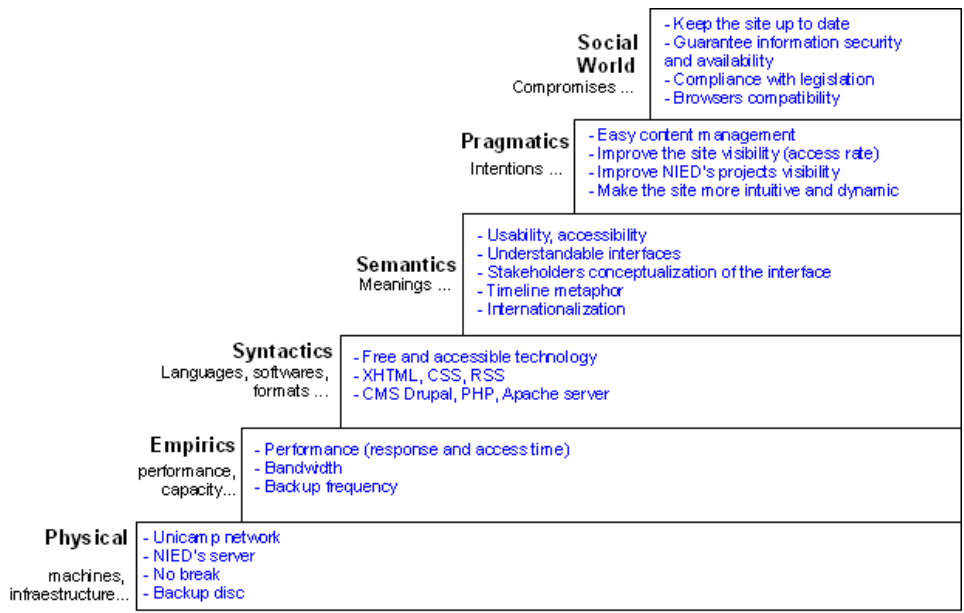

Figure 4. Semiotic ladder artifact applied to NIED's web portal.

### 5.3. Requirements Negotiation and Prototype

The second workshop results were discussed with the technical team and the coordination, aiming to delimit the scope of the (re)design.

The third participatory workshop was organized with the stakeholders for the purpose of negotiation. The requirements were classified into functional and non-functional.

Firstly, the semiotic ladder and the evaluation frame developed in the last workshop were revisited. Afterward, the functional and non-functional requirements were presented and considered one by one.

The most prioritized functional requirements were related to other ways to spread news (RSS and social networks), to ease the access to NIED's books and projects, to customize projects and staff pages, to decentralize content management, and to represent NIED's community.

The non-functional requirements highlighted by the stakeholders were related to accessibility, usability, metaphor to content organization (timeline view), visual identity, updated content, and emphasis on NIED's projects.

After some discussion, the stakeholders agreed that the following items should be reformulated based on the coordination recommendations and the technical team limitations:

- Electronic commerce does not fit in NIED's educational nature;
- There is already an Intranet for some administrative matters;
- Community contributions to projects and blogs also imply in responsibilities with posted content;
- Digital library and mobile devices access were considered interesting requirements, but due to the limited development team they will not be considered for now.

After the requirements negotiation, we started a participatory prototype in order to know what the stakeholders expected from the portal main page. The braindraw participatory technique [7] was adopted to this activity. The stakeholders were separated in pairs. Each pair had 5 to 10 minutes to draw the main page of the portal on a sheet. The sheet with the contributions of each pair was passed to the right until all pairs had contributed in each sheet.

The result was very interesting: part of the stakeholders focused on NIED's projects and the other part focused on the institution, which resulted in two prototypes for the main page. The prototypes were discussed with the stakeholders in the requirements validation phase (session 5.6).

### 5.4. Semantic Analysis

After the 3rd participatory workshop, the analyst and development team accomplished the Semantic Analysis in order to formalize and model the system requirements. The semantic analysis can be summarized in four main parts: problem definition, generation of candidate affordances, affordances grouping, and ontology diagram creation [4].

The computational problem definition was introduced in section 5.1 by the following statement: "the development of a web portal to meet NIED's interdisciplinary nature and to highlight the nucleus projects and advances". The problem statement can serve as an initial source of agents and affordances. But, like any initial statement, vague as this one, the information that resulted from the participatory workshops contributed much more to the discovery of agents, affordances, system properties, and so on. The candidate affordances are listed below.

```
    portal, site map, news, events, time line, team, contact, mission,
     vision, projects, publications, UCA, TelEduc, XO, FAQ, languages,
    address, phone, location, photos, videos, partnerships, RSS, logo,
     community, restricted access, creative-commons, papers, research,
      twitter, facebook, books, login, password, menus, lattes, memos,
  regulation, reports, user, W3C accessibility, participants, templates,
 social networks, translation, permissions, edit, responsible, upgrade.
```

During the next phase of semantic analysis – affordances grouping – new relevant semantic units were identified. The following affordances grouping refers to an overview of the technological context.

```
Agents:
Society (root agent)
  Person (agent - affordance of Society - part of Society)
    User (role of a Person in the portal)
      Administrator (User specification)
      Researcher (User specification)
      Teacher (User specification)
      Secretary (User specification)

  Internet (agent - affordance of Society - part of Society)
    Web (agent - affordance of Internet - part of Internet)
      Portal (agent - affordance of Web - part of Web)
        Administration (agent - affordance of Portal - part of Portal)
        Session (agent - affordance of Portal - part of Portal)
          Site Map (Portal specification)
          News (Portal specification)
          Events (Portal specification)
          Time line (Portal specification)
          Team (Portal specification)
            Lattes (Team property)
          Mission (Portal specification)
          Projects      (Portal specification)
          Publication (Portal specification)
            Memos (Publication specification)
            Books (Publication specification)
            Papers (Publication specification)
          FAQ (Portal specification)
          Location, Contact (Portal specification)
          Photos, Videos, Logo (Portal specification)
          Partnerships (Portal specification)
          RSS, twitter, facebook (Portal specification)

Affordances:
- Data (User property - determiner)
- Browse the portal (User affordance)
- Responsible for (User affordance)
- Language (Portal affordance)
- Accessibility links (Portal affordance)
- Choose (User affordance, depends on Language and Accessibility links)
- Identification (User affordance, depends on User Data)
- Edition (User affordance, depends on User Identification)
- Team (Portal affordance, depends on User Identification)
- Projects (Portal affordance, depends on User Identification)
```

```
- FAQ (Portal affordance, depends on User Identification)
- News (Portal affordance, depends on User Identification)
- Events (Portal affordance, depends on User Identification)
- Publications (Portal affordance, depends on User Identification)
- Partnerships (Portal affordance, depends on User Identification)
```

The affordances grouping presented in this section is a general view of NIED's web portal. A detailed description can be given for each affordance, using the same format. For example, the affordance "Events" may have properties "begin date", "begin time", "end date", "end time", "location", "duration", "summary", "related links", and so on.

As the last part of the semantic analysis, the ontology diagram was constructed using the identified agents and affordances. Figure 5 illustrates an ontology diagram that describes the domain problem as a whole. Specific parts of the problem, such as "News", "Publications", "Events", and others, can be represented in specific ontology diagrams. The diagram represents the understanding and the expression of the designer for NIED's web portal and its context of use. Such understanding takes into account the expectations of users who have worked in the participatory activities.

In Figure 5 diagram, there is a root agent – "Society" – where users' actions and meanings on the Internet are shared. The ontological dependence between agents and affordances is represented from left to right. For example, the affordance "browse" depends on the existence of a "User" – "Person" role in "Site" – and the affordance "language" depends on the existence of the "Site" agent, where this language is expressed.

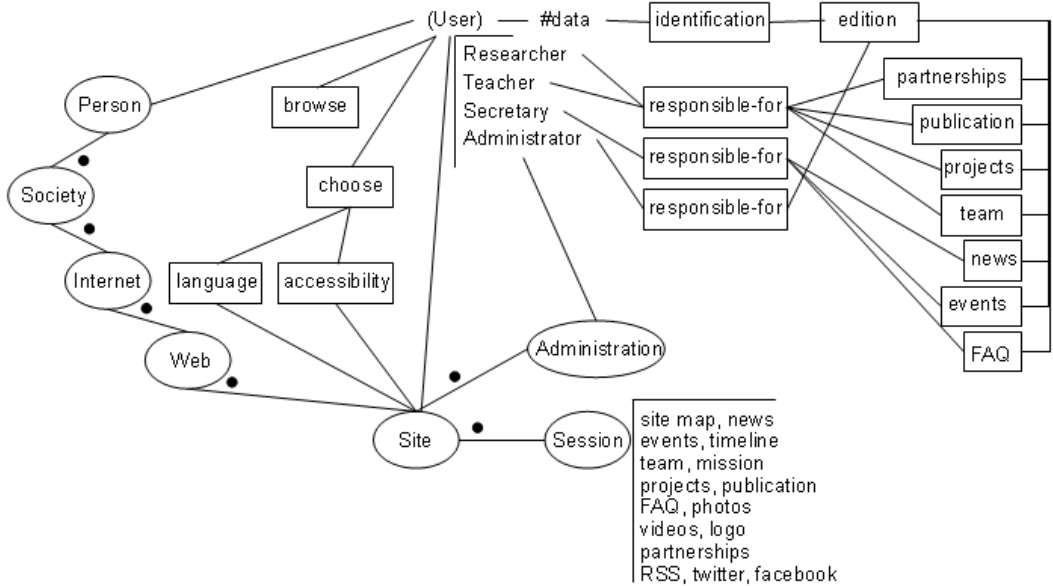

Figure 5. Ontology Diagram – problem context general view, based on the affordances grouping.

## 5.5. Requirements Specification

At this stage, the requirements specification document was produced. It contains the following sections:

- Introduction: The main topics are related to goals and target audience of the document, system scope, and document overview.
- General Description: It specifies user characteristics, system perspective, functions, constraints, assumptions, and dependencies. The general ontology diagram showed on Figure 5 was used to illustrate the system functions.
- Functional Requirements: The system requirements were described in natural language. The requirements related to the affordances "Edition", "Team", "Projects", "FAQ", "News", "Events", "Publications", and "Partnerships" were detailed in specific ontology diagrams.
- Non-Functional Requirements described in natural language.

## 5.6. Requirements Validation

The fourth and last workshop was conducted to make sure that the requirements document produced corresponds, in fact, to the system that the stakeholders intends to.

The requirements were reviewed one by one using natural language description and the ontologies diagrams.

The two prototypes of the portal main page, resulted from the 3rd workshop braindraw (section 5.3), were presented. One of them had its focus on NIED's projects, while the other one focused NIED as institution. As stakeholders continued in doubt about which one to choose, they suggested to join them into one, resulting in the main page prototype illustrated in Figure 6.

As shown in Figure 6, the portal prototype has five areas: the header containing logos, accessibility links, internationalization links, and main menu; the left area showing NIED's projects with photos, video and a brief description; the middle (or main) area containing NIED's description with links and photos related to news and events; the right area with a calendar, and links to news and events; and finally, the footer showing creative commons license, NIED's location, contact, RSS and social networks links.

## 6. DISCUSSION

It is an agreement that software cannot function in isolation from the organizational and social context in which it is embedded [2, 4, 15, 16]. The design of technological products is strongly related to social and organizational dynamics.

According to Erickson [15], the design process can be more effective by developing a better understanding of how concrete artifacts support communication in design. According to Kuutti [16] the organizational context where a computer system is embedded is a social system, so the relationship between the social system and the technical one should be considered and discussed in the system design [16].

OS artifacts in conjunction with PD techniques have been explored in previous work of this research group [9, 10, 11]. According to Baranauskas et al. [2], the OS methods allowed to encompass a system level view, involving the team into considerations about the formal and informal levels; instead of emphasizing the behavior of the software system as usually proposed by traditional methodologies.

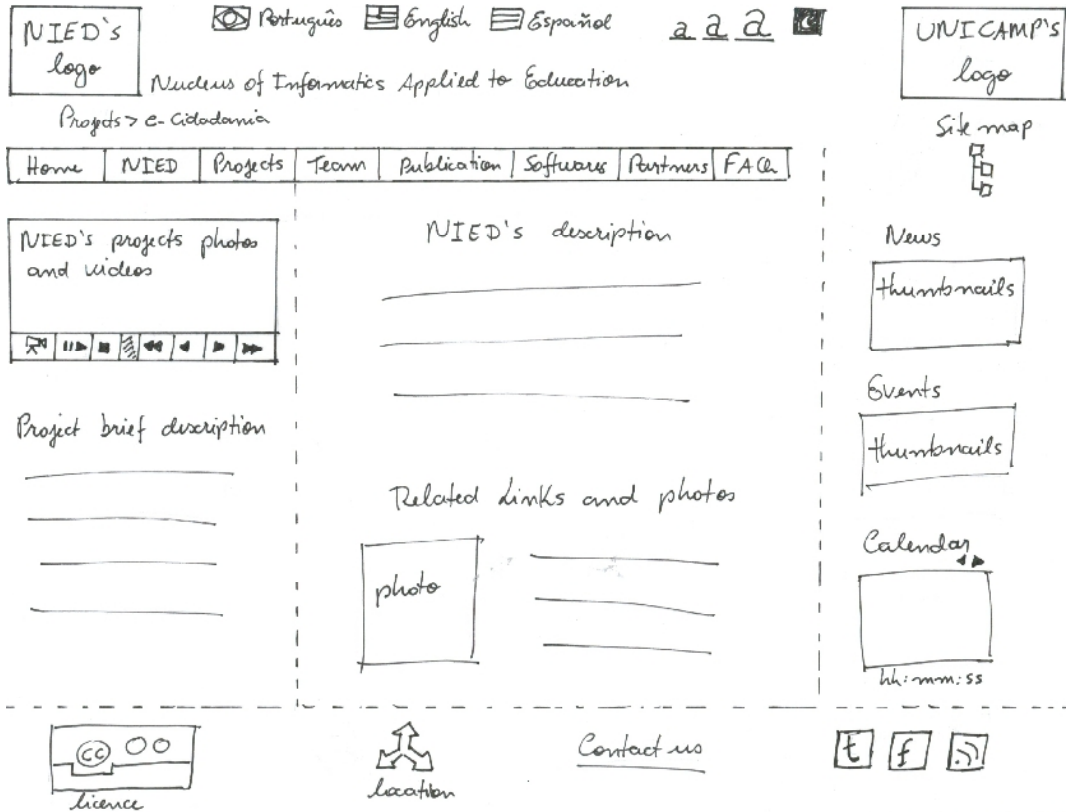

Figure 6. The main page prototype with focus on both - NIED's projects and NIED as institution.

The artifacts and activities carried out during the workshops deal mainly with the three upper layers of the semiotic ladder, which are concerned with the use of signs to communicate meanings (semantic layer), intentions (pragmatic layer), and their social consequences (social layer). The activities executed in the workshops dealt with information that involves cultural, behavioral and ethical aspects.

As a multidisciplinary and human-centred process, the participatory approach facilitated discussion among the stakeholders during the workshops, leading to a better understanding of the context and social implications of the system. Communication between the stakeholders was facilitated by the OS artifacts, which allowed meaning negotiation and context understanding by a shared representation of the data being captured, analyzed, and discussed. With participatory workshops the concepts were naturally originated from the stakeholders themselves, who express their ideas in post-its, and discuss their viewpoints and goals with others in the group.

Using the artifacts, a list of agreed requirements was derived, allowing a big picture of the problem and the main requirements. The requirements considered stakeholders' needs, intentions, beliefs, existing conflicts, etc. They were represented with the use of an ontology model, providing elements to specification and documentation.

## 7. CONCLUSION

Inefficient RE has been one of the causes of systems that do not meet their customers' requirements; like happened with NIED's current web portal when it was delivered.

In this paper, an experience with NIED's web portal was presented. It showed how the semio-participatory methodology [8] can be used to join OS artifacts and PD practices to support web sites Requirements Engineering. The outlined methodology, illustrated in Figure 2, can be considered a general guide to be applied in other web sites RE phases.

Results achieved in this case study, as well as in other application of the theoretical framework of OS to information system design [3, 9, 10], encourage its use in other information systems engineering.

The portal implementation is a work in progress. Some future work includes participatory evaluation and requirements management.

## ACKNOWLEDGEMENTS

Thanks to users that took part in the participatory workshops.

## .AUTHOR

Flávia Linhalis Arantes received the Ph.D. degree in Computer Science from Universidad e de São Paulo, São Carlos, Brazil, in 2007. She is currently a researcher at Nucleus of Informatics Applied to Education (NIED), State University of Campinas (UNICAMP), Campinas, Brazil. Her research interests include participatory design, social web and e-learning.

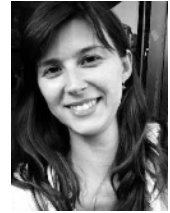